\documentclass[12pt]{iopart}

\usepackage{latexsym}
\usepackage{setstack}
\usepackage{graphicx}
\usepackage{iopams}
\usepackage{dsfont}
\usepackage{color}

\newcommand{\ket}[1]{\ensuremath{|#1\rangle}}

\newcommand{\be}{\begin{equation}}
\newcommand{\ee}{\end{equation}}
\newcommand{\mc}[1]{\ensuremath{\mathcal{#1}}}
\newcommand{\bc}{\begin{center}}
\newcommand{\ec}{\end{center}}

\newcommand{\ve}{\varepsilon}
\newcommand{\vro}{\varrho}
\newcommand{\mean}[1]{\ensuremath{ \langle #1  \rangle}}

\newcommand{\spl}{\ensuremath{ \mc{L}_{\mathrm{D}}}}
\newcommand{\op}{\ensuremath{\hat{a}}}
\newcommand{\opd}{\ensuremath{ \hat{a}^{\dagger}}}
\newcommand{\tc}{T_{\mathrm{loc}}}
\newcommand{\tp}{\tau_{\mathrm{c}}}
\newcommand{\np}{N_{\mathrm{ph}}}

\begin{document}

\title{Dissipation-induced correlations in 1D bosonic systems}

\author{Martin Kiffner and  Michael J Hartmann}

\address{Technische Universit\"at M\"unchen, Physik-Department I, 
James-Franck-Stra{\ss}e, 85748 Garching, Germany}

\begin{abstract}
The quantum dynamics of interacting bosons in a one-dimensional system 
is investigated numerically. 
We consider dissipative and conservative two-particle interactions, and 
integrate the   master equation describing the system dynamics   via a
time-evolving block-decimation (TEBD) algorithm. 
Our numerical simulations directly apply to  stationary-light polaritons 
in systems where atoms and photons are confined to the hollow 
core of a photonic crystal fibre. 
We show that  a two-particle loss term can drive  an initially uncorrelated state into 
a  regime where correlations effectively inhibit the dissipation of particles.
The  correlations induced by two-particle losses are compared  
with those generated by an elastic repulsion.  
For the considered time range, we find a similar behaviour in 
local density-density correlations but differences in non-local correlations. 
\end{abstract}

\pacs{05.30.Jp,03.65.Yz,42.50.Ex,42.50.Gy}

\submitto{\NJP}

\maketitle

\tableofcontents

%\cite{}
%
%\cite{baur:10}

\section{Introduction \label{intro}}
The preparation of quantum systems in entangled many-body states 
is an indispensable resource  for quantum information science~\cite{nielsen:00} 
and  allows one to access the rich physics of strongly correlated 
many-body systems~\cite{bloch:08}. 
Despite its reputation as the main adversary of quantum information schemes, 
dissipation was recognised recently as  a versatile tool for the generation of 
specific many-body states~\cite{kraus:08,verstraete:09,diehl:08}.

The preparation of correlated quantum states via dissipation requires that 
the target state itself is metastable. 
An intriguing experiment~\cite{syassen:08}  with cold molecules showed that  
a dissipative two-particle interaction between bosons in 1D 
gives rise to correlations that inhibit  the loss of particles. 
The formal analysis of this effect is based on a generalised  
Lieb-Liniger model~\cite{ripoll:08,duerr:09,baur:10}
and suggests that  the two-particle loss term 
effectively results in a repulsion such that inelastic collisions 
between particles are avoided. 
In the limit of strong  interactions~\cite{duerr:09}, the two-particle losses give rise  
to a Tonks-Girardeau gas~\cite{girardeau:60}  
where bosons behave with respect to many observables as if they were fermions. 
Recently, it was shown that 1D systems of interacting bosons can be realised 
via dark-state polaritons that arise in light-matter interactions under 
conditions of electromagnetically induced transparency~\cite{andre:05,HBP06,chang:08,kiffner:10,kiffner:10b}.  
A possible experimental realisation of the polariton schemes  
is  the setup described in~\cite{bajcsy:09,vorrath:10},  
where photons and atoms are simultaneously confined to the hollow core of a 
photonic-crystal fibre, see Fig.~\ref{fig1}.  
Another approach comprises 
the experimental setup in~\cite{vetsch:09}, where atoms are trapped near the 
surface of an optical nanofibre using evanescent waves. 
The nature of the polariton-polariton interaction in these approaches 
can be tuned by external parameters and 
can be elastic~\cite{chang:08} or dissipative~\cite{kiffner:10,kiffner:10b}, and 
the interaction strength is maximal for a purely dissipative interaction~\cite{kiffner:10}. 
Polariton schemes~\cite{chang:08,kiffner:10,kiffner:10b,hartmann:10,leib:10,hafezi:09a} 
are extremely promising candidates for the realisation of strongly correlated photon states. 

Here we consider bosons in 1D and investigate the quantum dynamics after a 
sudden switch-on of elastic and inelastic interactions 
via time-dependent density matrix renormalisation group techniques.
The master equation of the system and the uncorrelated initial state are chosen such that 
our simulations directly apply to polariton systems~\cite{chang:08,kiffner:10,kiffner:10b}.  
All details of our model are discussed in Sec.~\ref{model}, and 
numerical results are presented in Sec.~\ref{results}. 
In particular, we are interested in the preparation of a correlated, 
non-decaying state via dissipative two-particle interactions that 
we discuss in  Sec.~\ref{prep}. 
Non-local correlations are investigated in Sec.~\ref{comp}, where  we 
show that dissipation and repulsion give rise to  different results. 
Finally, a summary  is provided in Sec.~\ref{summary}. 
\section{Model \label{model}}
We consider bosons of mass $m$ in a one-dimensional system of length $L$ that 
experience a two-particle contact interaction that can be either 
conservative, dissipative or a mixture of both. 
First we describe the full master equation governing the 
time evolution of the system in Sec.~\ref{fm}. 
A physical realisation of this model via interacting 
dark-state polaritons~\cite{kiffner:10,kiffner:10b} is introduced in Sec.~\ref{phys}. 
Finally, we discuss the discretised model of the master equation in Sec.~\ref{fm} 
that allows us to perform numerical simulations in Sec.~\ref{disc}. 
\subsection{Master equation \label{fm}}
The  dynamics of bosons in our 1D system is described by the master equation
\be 
\hbar \partial_t \vro = -i H_{\mathrm{eff}}\vro +i \vro H_{\mathrm{eff}}^{\dagger} + \mc{I}\vro 
+ \hbar \spl  \vro , 
\label{meq}
\ee
where $H_{\mathrm{eff}}$ is a non-hermitian Hamiltonian, 
\be
H_{\mathrm{eff}} = \frac{\hbar^2}{2 m} \int_0^L \mathrm{d}z
 \partial_z\psi^{\dagger}  \partial_z\psi  
+  \frac{\tilde{g}}{2}\int_0^L \mathrm{d}z  \psi^{\dagger 2}  \psi^2. 
\label{Heff}
\ee
Here $\psi(z)$ is the field operator  obeying bosonic  commutation relations
\be
[\psi(z),\psi^{\dagger}(z^{\prime})]=\delta(z-z^{\prime}), 
\ee
and the first term in Eq.~(\ref{Heff}) describes the kinetic energy.  
The parameter $\tilde{g}$ is the complex coupling constant and 
its real part  gives 
rise to a hermitian contribution to $H_{\mathrm{eff}}$ in Eq.~(\ref{Heff}) that accounts for 
elastic two-particle collisions. 
On the other hand, the imaginary part of $\tilde{g}$ 
together with the term 
\be
\mc{I}\vro = - \mathrm{Im}(\tilde{g}) \int_0^L \mathrm{d}z \psi^2 \vro  \psi^{\dagger 2}
\label{ird} 
\ee
in Eq.~(\ref{meq}) constitutes a two-particle loss term 
that can be written 
in Lindblad form~\cite{breuer:os}  as  $\mathrm{Im}(\tilde{g}/2)\mc{D}[\psi^2]$, where 
$\mc{D}[\hat{X}]=\int_0^L \mathrm{d}z
(\hat{X}^{\dagger}\hat{X}\vro+\vro\hat{X}^{\dagger} \hat{X}-2\hat{X}\vro\hat{X}^{\dagger})$.   
The last term in Eq.~(\ref{meq})  is defined as 
\be
 \spl\vro= -\frac{D}{2}
\int_0^L \mathrm{d}z  \mc{D}[\partial_z\psi] 
\label{spl}
\ee
and causes diffusion of the field $\psi$.  
Throughout this paper we choose 
\be 
D= |\hbar/m|/10
\label{dval}
\ee
such that the losses associated with the diffusion term 
are small on a timescale set by  the kinetic energy term in Eq.~(\ref{Heff}). 

The master equation~(\ref{meq}) without the last term $\spl\vro$ 
defined in Eq.~(\ref{spl}) can be identified with the 
dissipative Lieb-Liniger model~\cite{duerr:09}. 
For a fixed number of particles $\np$, essential features of the 
system are described by the complex Lieb-Liniger parameter
\be
G = \frac{m \tilde{g}}{\hbar^2 \np/L}. 
\ee
The spectrum of the effective Hamiltonian in Eq.~(\ref{Heff}) was 
analysed in~\cite{duerr:09} via the Bethe-Ansatz. It was found 
that the ground state of the so-called gaseous states (all momenta 
converge to finite values for $|G|\rightarrow\infty$) approaches 
the Tonks-Girardeau gas in the limit $|G|\rightarrow\infty$. 
This state is metastable since in the Tonks-Girardeau gas limit 
two particles never occupy the same position in space, and hence 
the two-particle contact interaction does not contribute to the 
dissipation of particles. 

This feature can be formally derived from the master equation~(\ref{meq}). 
The two-particle loss term $\mathrm{Im}(\tilde{g}/2)\mc{D}[\psi^2]$ in Eq.~(\ref{meq}) 
causes a time dependence of the density of particles according to 
\be
\partial_t\mean{\hat{n}(z)} = \frac{2}{\hbar} \mathrm{Im}(\tilde{g})
 g^{(2)}(z,z) \mean{\hat{n}(z)}^2 , 
 \label{ntime}
\ee
where $\hat{n}(z)=\psi^{\dagger}(z)\psi(z)$ is the particle density operator and 
\be
g^{(2)}(z,z^{\prime})=
\frac{\langle\psi^{\dagger}(z)\psi^{\dagger}(z^{\prime})\psi(z)\psi(z^{\prime})\rangle}
{\langle \hat{n}(z)\rangle \langle \hat{n}(z^{\prime})\rangle}  
\ee
is the  second order correlation function. 
Note that the kinetic energy and the diffusion term $\spl\vro$ 
in Eq.~(\ref{meq}) will, in general, induce an additional time dependence of $\mean{\hat{n}(z)}$.  
The kinetic energy term conserves the total number of particles but gives rise 
to polariton fluxes from and to position $z$ that would redistribute the polariton density.
In regions where the system and the particle density are homogeneous, these fluxes can safely be neglected.  
The impact of the diffusion term $\spl\vro$ is negligible for the 
considered time scales since we chose a small 
value for the parameter $D$ in Eq. (\ref{dval}). 
In the strongly correlated regime and for a homogeneous system, 
the gaseous ground state of $H_{\mathrm{eff}}$ obeys~\cite{duerr:09}
\be
 g^{(2)}(z,z)= (1-1/\np^2)4\pi^2/(3 |G|^2), 
\label{smallg2}
\ee
and hence $g^{(2)}(z,z)\ll 1$ for $|G|$ sufficiently large. 
The combination of Eqs.~(\ref{ntime}) and~(\ref{smallg2}) implies 
that a purely dissipative interaction term supports 
metastable states. 
\subsection{Physical realisation \label{phys}}
%
%%%%%%%%%%%%%%%%%%%%
\begin{figure}[t!]
\bc
\includegraphics[width=14cm]{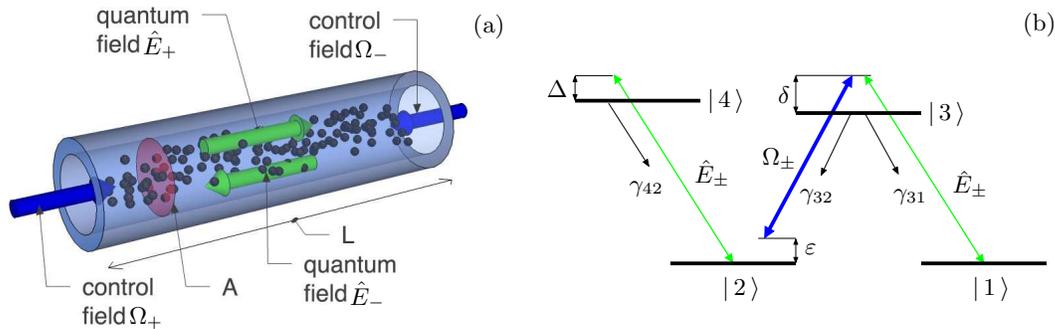}
\ec
\caption{\label{fig1} 
(a) Considered setup of 
$N_{\mathrm{a}}$ atoms   confined to an interaction volume of length $L$ and transverse area $A$. 
$\Omega_{\pm}$ are the Rabi frequencies of the classical control fields, and 
$\hat{E}_{\pm}$  are the quantum probe fields. 
(b) Atomic level scheme. $\gamma_{ij}$ is the full decay rate 
on the $\ket{i}\leftrightarrow\ket{j}$ transition, $\delta$ and $\Delta$ 
label the detuning of the probe fields with states $\ket{3}$ and $\ket{4}$, 
respectively, and $\ve$ is the two-photon detuning. 
}

\end{figure}
%%%%%%%%%%%%%%%%%%%%
%
% 
An example for a physical system that is described 
by the master equation~(\ref{meq}) is shown in Fig.~\ref{fig1}(a),  
where photons and atoms are simultaneously confined to the hollow core of a 
photonic-crystal fibre. 
Since the light-guiding core of the optical fibre is of the same order of magnitude 
as the optical wavelength, the fibre represents a one-dimensional waveguide 
for the optical fields. 
The level scheme of each atom inside the fibre is shown in Fig.~\ref{fig1}(b). 
The dynamics of the system can be described in terms of 
dark-state polaritons~\cite{fleischhauer:00} representing   
bosonic quasi-particles that arise in light-matter interactions under conditions of 
electromagnetically induced transparency (EIT)~\cite{fleischhauer:05}. 
The two counter-propagating control fields $\Omega_{\pm}$ [see Fig.~\ref{fig1}(a)] 
of equal intensity give rise to stationary 
light~\cite{andre:02,bajcsy:03} 
where the polaritons experience a quadratic dispersion relation like bosons in free space.  
Moreover, the coupling of the probe fields to the transition $\ket{2}\leftrightarrow\ket{4}$ 
induces a two-particle contact 
interaction between the polaritons that can be conservative~\cite{andre:05,chang:08},  
dissipative or a mixture of both~\cite{kiffner:10,kiffner:10b}. 
It can be shown~\cite{kiffner:10,kiffner:10b} that the dark-state polariton dynamics can 
be described by the master equation~(\ref{meq}), where the effective mass $m$ 
and the complex coupling constant $\tilde{g}$ depend on the detunings 
$\delta$, $\Delta$ [see Fig.~\ref{fig1}(b)] and the intensity of 
the control fields. 
In particular, we point out that a proper adjustment of the detuning $\delta$ 
allows one to fulfil Eq.~(\ref{dval}), and the choice of the detuning 
$\Delta$ determines the relative contribution of elastic and inelastic 
processes to the two-particle interaction. 

In an experiment, the evolution of the polaritons under Eq.~(\ref{meq}) has to be preceded by 
a loading process that could be realised as follows~\cite{chang:08}. 
Initially the probe field modes are empty and only the control 
field $\Omega_+$ is switched on. 
Then a  probe pulse copropagating with $\Omega_+$ enters the fibre under conditions 
of electromagnetically induced transparency.  
Here we assume that the transition $\ket{4}\leftrightarrow\ket{2}$ is far off-resonant 
during the loading process such that its influence  is negligible.  
The dynamics of the probe field inside the medium can be described concisely in terms 
of dark-state polaritons~\cite{fleischhauer:00}. 
As compared to the pulse propagating in vacuum, the polariton pulse inside 
the medium is spatially compressed and travels with a reduced group velocity  
$v_g\propto \Omega_+^2$. 
Once the pulse is entirely inside the fibre, 
the group velocity of the polariton pulse  is 
reduced to zero by an adiabatic switch-off of the control field $\Omega_+$. 
The latter process maps the polaritons into a stationary spin coherence, 
and all probe field modes are empty. 
Eventually both control fields  are switched on adiabatically and 
with equal intensity, $\Omega_{\pm}=\Omega_0$. 
The time evolution of the corresponding dark-state polaritons is then 
determined by the master equation~(\ref{meq})~\cite{kiffner:10,kiffner:10b}. 
Note that the adiabatic switch-on can be much faster than all typical timescales 
in the master equation~(\ref{meq}). 
At this stage, the detuning $\Delta$ of the probe field with the $\ket{4}\leftrightarrow\ket{2}$ 
transition can be adjusted by a proper choice of the frequency of 
the counterpropagating control fields. This frequency can be different from 
the one used for the single control field during the loading process.

The correlations that built up 
under the evolution of the master equation~(\ref{meq}) can be measured 
if the intensity of the control field $\Omega_-$ is reduced. 
The latter process  converts the stationary polaritons into 
a propagating pulse with controllable group velocity. 
This procedure maps spatial correlations of the trapped pulse 
into temporal correlations of the output 
light that can  be detected via standard quantum optical techniques. 
\subsection{Discretised model \label{disc}}
The dynamics of the system is studied via the time-evolving block decimation 
algorithm~\cite{zwolak:04,verstraete:04,hartmann:09} for 
density operators. 
The discretised version~\cite{schmidt:05} of Eq.~(\ref{meq}) is given by a generalised 
Bose-Hubbard model augmented by loss terms,
\be 
 \dot{\vro} = -\frac{i}{\hbar} [H_1 + H_2,\vro] + \mc{L}_1  \vro  + \mc{L}_2\vro .
\label{meq2}
\ee
Here $H_1$ and $H_2$  correspond to the kinetic energy term and the conservative 
contribution to the two-particle interaction, respectively,
\be 
 H_1 = 2 J \sum\limits_l \opd_l\op_l -J\sum\limits_l (\op_{l}\opd_{l+1} + \op_{l+1} \opd_{l}), \qquad
 H_2 = \frac{U}{2} \sum\limits_l \op_l^{\dagger 2} \op_l^2,
\ee
and $J$ and $U$ are defined as 
\be
 J = \frac{\hbar^2}{2 m}\frac{1}{\Delta z^2},  \qquad
U  = \mathrm{Re}(\tilde{g}) \frac{1}{\Delta z}. \\
\ee
The parameter $\Delta z$ is the lattice constant of the discretisation grid. 
The terms $\mc{L}_1\vro$ and $\mc{L}_2\vro$ in Eq.~(\ref{meq}) are the discretised versions of 
the diffusion term in Eq.~(\ref{spl}) and the dissipative contribution to the two-particle 
interaction, respectively,
\begin{eqnarray}
 \mc{L}_1  \vro = & -  \Gamma_1   \sum\limits_l 
(\op_l^{\dagger} \op_l \vro +\vro \op_l^{\dagger} \op_l -2 \op_l \vro \op_l^{\dagger} ) \\
& + \frac{\Gamma_1}{2}  \sum\limits_l 
(\op_l^{\dagger} \op_{l+1} \vro +\vro \op_l^{\dagger} \op_{l+1} -2 \op_{l+1} \vro \op_l^{\dagger} ) \nonumber \\
& + \frac{\Gamma_1}{2}  \sum\limits_l 
( \op_l \op_{l+1}^{\dagger} \vro +\vro \op_l \op_{l+1}^{\dagger} -2 \op_{l} \vro \op_{l+1}^{\dagger} ), \nonumber 
\end{eqnarray}
\be
 \mc{L}_2\vro = - \frac{\Gamma_2}{2} \sum\limits_l 
(\op_l^{\dagger 2} \op_l^2 \vro +\vro \op_l^{\dagger 2} \op_l^2 -2 \op_l^2 \vro \op_l^{\dagger 2} ), 
\ee
and 
\be
\Gamma_1 = D\frac{1}{\Delta z^2}, \qquad
 \Gamma_2 = - \mathrm{Im}(\tilde{g}/\hbar)\frac{1}{\Delta z} .
\ee
The discretised version Eq.~(\ref{meq2}) is a good approximation 
of the continuous master equation~(\ref{meq})  if the smallest wavelength 
$\lambda_{\mathrm{min}}$ involved is much larger than the grid spacing $\Delta z$. 
Initially $\lambda_{\mathrm{min}}$ is determined by the momentum components of 
the initial state, but the evolution under Eq.~(\ref{meq2}) will 
change the momentum distribution of the sample. 
In particular, the suddenly switched-on two-particle repulsion 
term  leads (among other processes) to a population of higher momentum states. 
It was found~\cite{muth:10}  that the lattice model is a good approximation 
provided that $U/J\ll 1$.  
If the latter inequality is violated, lattice artefacts occur. 
We find that these effects do not occur for a purely dissipative 
two-particle interaction ($U=0$) even if  $\hbar\Gamma_2 /J \approx 1$. 

For all numerical simulations we consider a grid 
with $N_{\mathrm{s}}=500$ sites, and the physical state space at each 
site $l$ is spanned by the four Fock states 
$\{\ket{0}_l,\ket{1}_l,\ket{2}_l,\ket{3}_l\}$ allowing for 
a maximum occupation of three particles at each site.   
We verified the convergence of our algorithm by varying the 
bond dimension $\chi$ of the matrix product state. 
In our simulations, the truncation errors $\epsilon_{\mathrm{trunc}}$ 
are bounded by $\epsilon_{\mathrm{trunc}}\le 10^{-4}$ for $\chi=100$. 
%

%
%%%%%%%%%%%%%%%%%%%%
\begin{figure}[t!]
\bc
\includegraphics[scale=1]{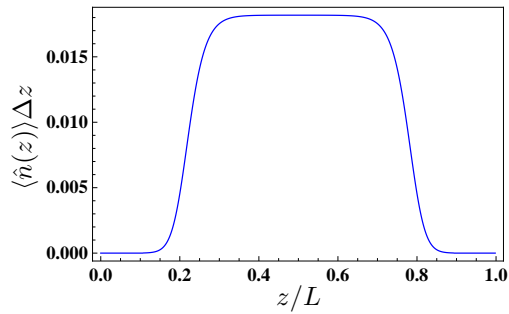}
\ec
\caption{\label{fig2} 
Spatial density distribution $\mean{\hat{n}(z)}$ 
of the initial state in Eq.~(\ref{init}). 
}
\end{figure}
%%%%%%%%%%%%%%%%%%%% 
%
The choice of the initial state for the numerical integration is 
motivated by the loading process of an optical pulse into the fibre  as 
described in Sec.~\ref{phys}.  
We consider an initial state with a mean number of $\mean{\hat{N}}_0=5$ particles and 
a spatial density distribution as shown in Fig.~\ref{fig2}. 
Here we neglect distortions and correlations due to the optical nonlinearity 
that may build up during the loading process.  
The initial shape of the polariton pulse is then determined by the 
slowly varying envelope of the  probe pulse entering the 
system~\cite{fleischhauer:00}. 
Furthermore, we assume that the initial polariton pulse is derived from  laser light 
and thus described by a product of coherent states. 
In the discretised model we thus model the initial state by a coherent 
state at each site, 
\be
 \ket{\psi}_0 = \prod\limits_{l} \exp(-|c_l|^2/2)\exp(c_l a_l^{\dagger})\ket{0} , 
 \label{init}
\ee
where $\sum_l |c_l|^2 = 5$ and   $|c_l|^2= \Delta z \mean{\hat{n}(z_l)}$ 
with $z_l = l \Delta z$. 
We point out that $\ket{\psi}_0$  in Eq.~(\ref{init}) is an uncorrelated state with 
$g^{(2)}(z,z^{\prime}) = 1$ for all values of $z,z^{\prime}$ in the sample. 
\section{Numerical results \label{results}}
%
%%%%%%%%%%%%%%%%%%%%
\begin{figure}[t!]
\bc
\includegraphics[width=12.5cm]{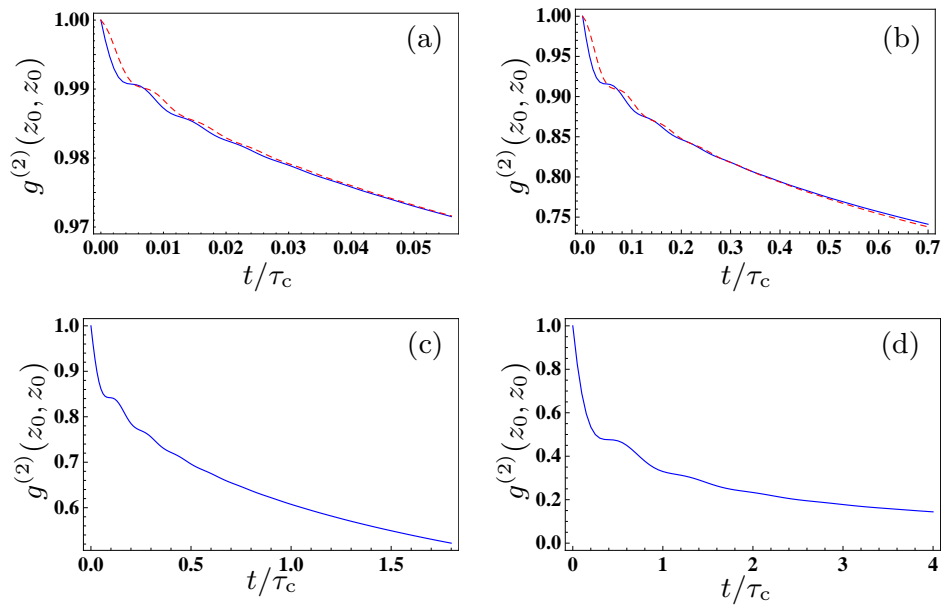}
\ec
\caption{\label{fig3} 
Temporal evolution of $g^{(2)}(z_0,z_0)$ 
at the centre $z_0=L/2$ of the sample for (a) $|G|=1$, 
(b) $|G|=10$, (c) $|G|=20$ and (d) $|G|=100$.  
The timescale $\tp$ is defined in Eq.~(\ref{tp}), and 
the solid lines correspond to a purely dissipative two-particle 
interaction, $\mathrm{Re}(\tilde{g})=0$. 
In (a) and (b), the dashed line shows the result for 
an interaction that is predominantly 
repulsive [$|\mathrm{Re}(\tilde{g})/\mathrm{Im}(\tilde{g})|=10$]. 
}
\end{figure}
%%%%%%%%%%%%%%%%%%%%
%
Next we describe the results of the numerical integration of 
the master equation~(\ref{meq2}). Starting from the initial state 
in Eq.~(\ref{init}), the interaction is suddenly switched on at $t=0$.  
We consider various scenarios where the interaction is purely 
dissipative or predominantly conservative.  
In Sec.~\ref{prep}, we discuss whether two-particle losses alone are able 
to drive the system into a correlated regime where dissipation of 
particles is suppressed. 
As discussed in Sec.~\ref{fm}, the inhibition of two-particle losses 
is linked to local correlations in the sample.
The second part Sec.~\ref{comp} is concerned with non-local 
correlations. In particular, we compare non-local 
correlations created by a purely dissipative  
and a predominantly conservative interaction. 
\subsection{Inhibition of two-particle losses \label{prep}}
The two-particle decay term in Eq.~(\ref{meq}) in general leads 
to a loss of particles from the system. 
The  rate at which the particle density diminishes is 
determined by Eq.~(\ref{ntime}) and depends  crucially on the 
local correlations $g^{(2)}(z,z)$ of  the system. 
Figure~\ref{fig3} shows the temporal evolution of $g^{(2)}(z_0,z_0)$ 
at the centre $z_0=L/2$ of the sample for different values of 
the Lieb-Liniger parameter $|G|$ according to the numerical integration 
of Eq.~(\ref{meq2}). The values of $|G|$ correspond to  
the mean particle number $\mean{\hat{N}}_0=5$ of the initial state, see Eq.~(\ref{init}).   
All solid lined lines in Fig.~\ref{fig3} 
belong to a purely dissipative two-particle interaction, $\mathrm{Re}(\tilde{g})=0$. 
On the contrary, the dashed lines in Fig.~\ref{fig3}~(a) and~(b) correspond 
to a predominantly conservative interaction 
$|\mathrm{Re}(\tilde{g})/\mathrm{Im}(\tilde{g})|=10$. 

In the case of a purely conservative interaction ($\mathrm{Im}(\tilde{g})=0$), 
it was shown~\cite{muth:10} 
that local quantities reach a stationary state on a timescale $\tc=\hbar/(\tilde{g} \rho)$, 
where $\rho=\np/L$ is the density of particles. 
Note that the Lieb-Liniger gas does not thermalize globally~\cite{kinoshita:06}. 
Here we generalise this definition to the case of complex $\tilde{g}$ and define 
\be
\tc = \frac{\hbar}{|\tilde{g}| \rho}=\frac{m}{\hbar}\frac{1}{|G|\rho^2} .
\label{tc}
\ee
Due to the enhanced numerical complexity of TEBD for density operators and 
numerical limitations,  our simulations cannot cover the full time range $\tc$. 
In the following we scale  time in the presentation of  numerical results with 
\be
 \tp= 2 \frac{\hbar}{\sqrt{U^2 +(\hbar\Gamma_2)^2}} =  \frac{1}{50} \tc. 
 \label{tp}
\ee
The comparison between the dashed and solid lines in Figs.~\ref{fig3}~(a) 
and~(b) shows that dissipative and repulsive  
two-particle interactions are equally effective in the repulsion of 
particles  on a timescale proportional to the inverse 
interaction strength $|\tilde{g}|$.  
Note that for the polariton system discussed in Sec.~\ref{phys}, 
$|\tilde{g}|$ is at least one magnitude larger  for a purely dissipative 
interaction~\cite{kiffner:10,kiffner:10b} as compared to the conservative case. 
It follows that small values of $g^{(2)}(z_0,z_0)$ are reached much faster 
in absolute time for a dissipative two-particle interaction. 

%
%%%%%%%%%%%%%%%%%%%%
\begin{figure}[t!]
\bc
\includegraphics[width=12.5cm]{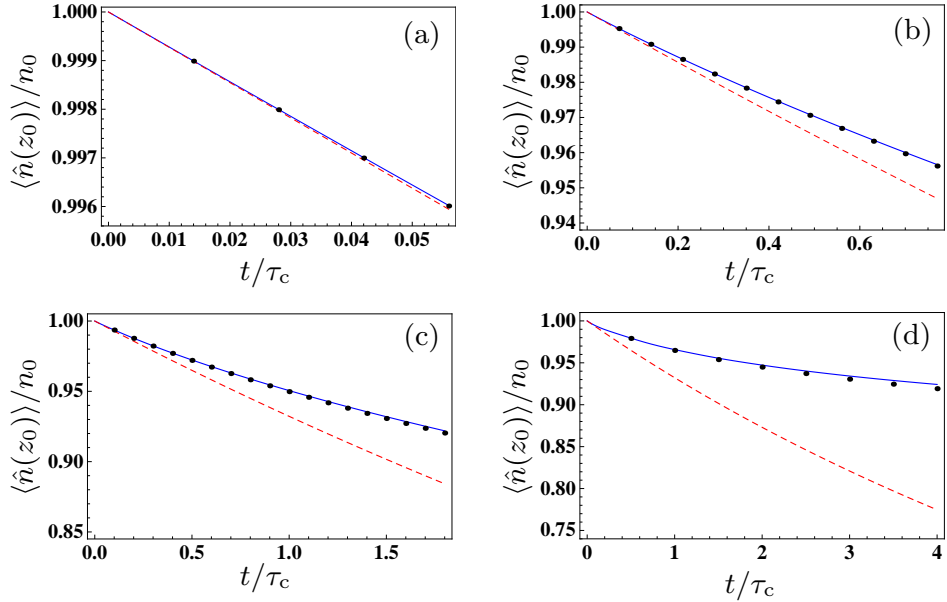}
\ec
\caption{\label{fig4} 
Temporal evolution of $\mean{\hat{n}(z)}$ 
at the centre $z_0=L/2$ of the sample for (a) $|G|=1$, 
(b) $|G|=10$, (c) $|G|=20$, (d) $|G|=100$.  
In (a)-(d), the solid line corresponds to a numerical integration 
of Eq.~(\ref{ntime}) with $g^{(2)}(z_0,z_0)$ taken 
from the TEBD simulation for a purely dissipative two-particle 
interaction, $\mathrm{Re}(\tilde{g})=0$. 
The dotted lines indicate the results of the TEBD simulation 
for $\mean{\hat{n}(z)}$. 
The dashed line shows the evolution of 
$\mean{\hat{n}(z)}$ according to Eq.~(\ref{ntime}) with 
$g^{(2)}(z_0,z_0)=1$  for all times.
}
\end{figure}
%%%%%%%%%%%%%%%%%%%%
%
We emphasise that the minimal values for $g^{(2)}(z_0,z_0)$ in Fig.~\ref{fig3} 
will decrease further in time until their  equilibrium values are reached. 
An estimate of these equilibrium values can be obtained via the 
work presented in~\cite{muth:10}. Here the authors performed TEBD simulations for the 
unitary time evolution  of bosons in 1D with two-particle repulsion. 
In particular, this system could be  described by a pure quantum state which 
reduces the numerical complexity as compared to our simulations of the 
full density operator. Therefore, the simulations in~\cite{muth:10} 
could be carried out up to times $t=\tc$. 
Given that the initial dynamics of $g^{(2)}(z_0,z_0)$ for two-particle 
losses and for conservative interactions is very similar, one 
may suppose that the equilibrium values of $g^{(2)}(z_0,z_0)$ will also be almost the same. 
For this case we estimate that $g^{(2)}(z_0,z_0)$ for our system 
should lie within the range $[0.1,0.2]$ for $|G|=10$ and $|G|=20$ at $t=\tc$. 

Next we determine the temporal evolution of the particle density 
at the centre of the cloud via Eq.~(\ref{ntime}) and the numerical 
results for the temporal evolution of $g^{(2)}(z_0,z_0)$. 
The result is represented by the solid lines  in Fig.~\ref{fig4} 
and in good agreement with the direct evaluation of the particle density $\mean{\hat{n}(z)}$ 
via the numerical integration of the master equation~(\ref{meq2}) (dotted lines).  
The small deviations are due to the diffusion term in Eq.~(\ref{spl})  that results in 
particle losses and that were omitted in Eq.~(\ref{ntime}). 
Note that all curves in Fig.~\ref{fig4} correspond to a 
purely dissipative two-particle interaction, $\mathrm{Re}(\tilde{g})=0$. 
In order to visualise the slowdown of two-particle losses due to 
the decrease of $g^{(2)}(z_0,z_0)$ in time, the dashed line in Fig.~\ref{fig4} 
shows the temporal evolution of $\mean{\hat{n}(z)}$ according to 
Eq.~(\ref{ntime}) with $g^{(2)}(z_0,z_0)=1$ for all times. 
The slowdown of two-particle losses is most pronounced for $|G|=100$ 
where a  quasi-stationary regime is reached with 
$g^{(2)}(z_0,z_0)$ close to zero. 
We find that the decrease of the mean number of particles $\mean{\hat{N}}/\mean{\hat{N}}_0$ 
follows the same curves as the particle density at the centre of the cloud. 
It follows that more than $90\%$ of the initially present  particles are left 
at $t/\tp=4$ for $|G|=100$. 
Figures~\ref{fig4}~(b) and~(c) correspond to   $|G|=10$ and $|G|=20$, respectively, 
and demonstrate a clear slowdown of two-particle losses. 
On the basis of the estimates for the equilibrium values of  $g^{(2)}(z_0,z_0)$ 
obtained via the results presented in~\cite{muth:10}, we suppose that  $50\%$ ($60\%$)  
of the initial particles are still present at $t=\tc$ for $|G|=10$ ($|G|=20$). 
On the contrary, $g^{(2)}(z_0,z_0)$ will not drop to values close to zero for  $|G|=1$ 
and hence we expect that a significant number of particles is lost at $t=\tc$. 
\subsection{Non-local correlations  \label{comp}}
%
%%%%%%%%%%%%%%%%%%%%
\begin{figure}[t!]
\bc
\includegraphics[width=12.5cm]{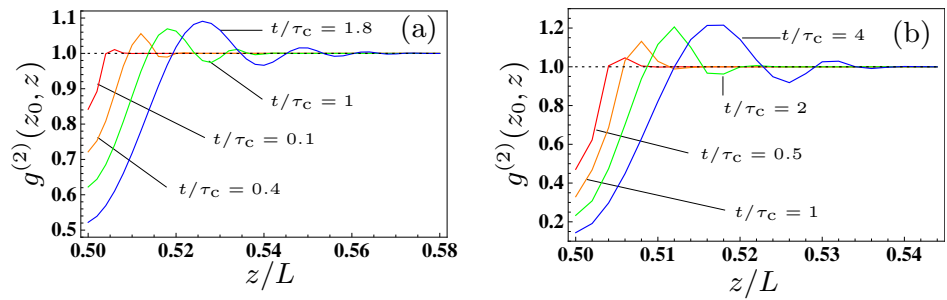}
\ec
\caption{\label{fig5} 
Spatial dependence of  $g^{(2)}(z_0,z)$ 
for a purely dissipative two-particle 
interaction [$\mathrm{Re}(\tilde{g})=0$]. 
The parameters correspond to (a) $|G|=20$ and 
(b) $|G|=100$, and $z_0=L/2$ is the centre of the sample.  
}
\end{figure}
\begin{figure}[t!]
\bc
\includegraphics[width=12.5cm]{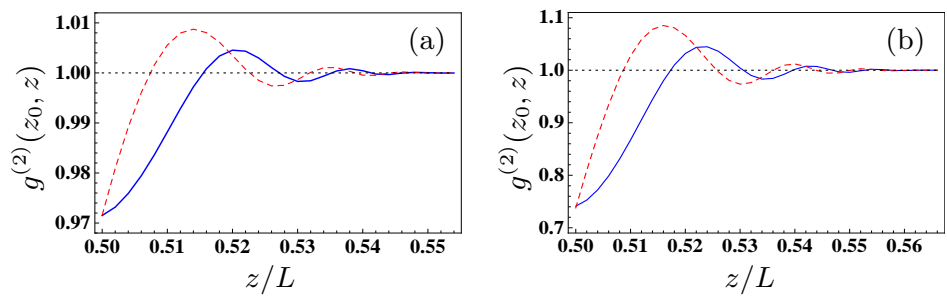}
\ec
\caption{\label{fig6} 
Spatial dependence of  $g^{(2)}(z_0,z)$ 
 for (a) $|G|=1$, $t/\tau_c=0.056$ and 
(b) $|G|=10$, $t/\tau_c=0.77$. $z_0=L/2$ is the centre of the sample. 
The solid lines correspond to a purely dissipative two-particle 
interaction  [$\mathrm{Re}(\tilde{g})=0$], and 
the dashed lines show the result for 
an interaction that is predominantly 
repulsive [$|\mathrm{Re}(\tilde{g})/\mathrm{Im}(\tilde{g})|=10$]. 
}
\end{figure}

%%%%%%%%%%%%%%%%%%%%
%
The spatial dependence of $g^{(2)}(z_0,z)$ as a function of $z$ 
($z_0=L/2$ is the centre of the sample) 
is depicted in Fig.~\ref{fig5}. 
Several snapshots in time are shown indicating that 
non-local correlations exhibit 
oscillations that propagate outwards. 
The oscillations for the longest evolution times in Fig.~\ref{fig5} 
exhibit several minima that lie below unity. 
It follows that a dissipative two-particle interaction alone 
to some extend gives rise to spatial order of the particles. 
Note that non-local correlations reach a quasi-stationary regime 
on a timescale that is much larger than $\tc$. 
In particular, we point out that the correlations   
shown in Fig.~\ref{fig6} 
are not Friedel oscillations~\cite{bloch:08} that characterise 
the Fermionic nature of particles in the Tonks-Girardeau regime. 

In Sec.~\ref{prep} we concluded that repulsion and dissipation are 
equally effective in the creation of small values of $g^{(2)}(z,z)$. 
On the other hand, dissipative and conservative two-particle interactions 
give rise to different non-local correlations as shown in Fig.~\ref{fig6}. 
The solid and dashed lines correspond to a purely dissipative and 
a mostly conservative two-particle interaction, respectively.   
It can be seen that the non-local correlations propagate at the same 
velocity in space in the  dissipative and conservative case, but 
the oscillations are more pronounced for a repulsive two-particle interaction. 
In addition, two-particle losses give rise to a broader dip 
in $g^{(2)}(z_0,z)$ as a function of $z$ as compared to the conservative case. 
Note that the oscillation amplitudes in $g^{(2)}(z_0,z)$ are also damped by 
the diffusion term in Eq.~(\ref{spl}). 
%
%%%%%%%%%%%%%%%%%%%%
%%%%%%%%%%%%%%%%%%%%
%
\section{Discussion and summary \label{summary}}
In this paper we  investigated the quantum dynamics of bosons in 1D 
after a sudden switch-on of two-particle losses. 
The master equation and the uncorrelated initial state 
are chosen such that our simulations directly 
apply to stationary-light polariton systems~\cite{chang:08,kiffner:10,kiffner:10b}.  
Apart from a (small) diffusion term, the considered master equation coincides with 
the dissipative Lieb-Liniger model~\cite{duerr:09}. 
In particular, we investigated the dissipation-induced preparation of  
correlations that effectively inhibit two-particle losses. 
We found that a metastable regime where losses are substantially suppressed 
can be prepared for large values of the Lieb-Liniger parameter $|G|\gg 1$. 
Formally, it was shown~\cite{ripoll:08,duerr:09} that  the two-particle loss term 
effectively results in a repulsion between particles such that they never occupy the 
same position in space. Therefore,  dissipation via the two-particle contact interaction 
is inhibited. A physical explanation for this counterintuitive result can be 
given for a  discrete lattice model~\cite{syassen:08,ripoll:08}. 
There it can be regarded as a manifestation of the quantum Zeno effect, 
where the two-particle losses play the role of a continuous measurement that 
always projects the system onto states with less 
than two particles in each lattice site before higher particle numbers can build up. 
Furthermore, we compared the correlations induced by dissipative 
and repulsive interactions, respectively. We found 
that dissipation and repulsion are equally efficient for 
the generation of local correlations $g^{(2)}(z,z)\le 1$ on a 
timescale proportional to the inverse interaction strength. 
On the other hand, non-local correlations induced by dissipation 
show  different features than those created by two-particle repulsion. 
As compared to an elastic repulsion between the particles, 
we find that two-particle losses give rise to a broader dip 
in $g^{(2)}(z_0,z)$ as a function of $z$ ($z_0$ is the centre of the sample).  
Both repulsion and dissipation give rise to  
spatial oscillations in $g^{(2)}(z_0,z)$ that spread in time 
and indicate to some extend a spatial ordering. 
We find that these oscillations are  more pronounced for  
a repulsion between the particles as compared to the dissipative case. 

% outlook

%%%%%%%%%%%%%%%%%%%%%%%%%%%%%%%%%%%%%%%%%%%%%%%%%%

\ack
The authors thank D. Muth and M. Fleischhauer for discussions. 
This work is part of the Emmy Noether project 
HA 5593/1-1 funded by the German Research Foundation (DFG).

\end{document}